\documentclass[aps,preprint,nofootinbib,showpacs]{revtex4}
\usepackage{graphics}
\usepackage{slashed}
\usepackage{amssymb}
\usepackage{lscape}
\usepackage{amsmath}
\usepackage{rotating}
\usepackage{graphicx}
\graphicspath{ {images/} }
\begin{document}

\title{Quantum thermodynamics in the interior of a Schwarzschild B-H}
\author{$^{2}$ Juan Ignacio Musmarra\footnote{jmusmarra@mdp.edu.ar}, $^{1,2}$ Mauricio Bellini\footnote{mbellini@mdp.edu.ar},  $^{1,2}$ Mariano Anabitarte\footnote{anabitar@mdp.edu.ar} }
\affiliation{$^1$ Departamento de F\'isica, Facultad de Ciencias Exactas y
Naturales, Universidad Nacional de Mar del Plata, Funes 3350, C.P.
7600, Mar del Plata, Argentina.\\
$^2$ Instituto de Investigaciones F\'{\i}sicas de Mar del Plata (IFIMAR), \\
Consejo Nacional de Investigaciones Cient\'ificas y T\'ecnicas
(CONICET), Mar del Plata, Argentina.}

\begin{abstract}
We study the interior of a Schwarzschild Black-Hole (B-H) using Relativistic Quantum Geometry described in \cite{rb} and \cite{rb1}. We found discrete energy levels for a scalar field from a polynomial condition for Heun Confluent functions expanded around the Schwarzschild radius. From the solutions it is obtained that the uncertainty principle is valid for each energy level of space-time, in the form: $E_n\, r_{sh,n}=\hbar/2$. Temperature, entropy and the B-H mass are dependent on the number of states in the B-H, such that the Bekenstein-Hawking (BH) results are obtained in a limit case.
\end{abstract}
\maketitle

\section{Introduction and motivation}

The Schwarzschild solution for a B-H is very important for conceptual discussion in general relativity, in particular regarding event horizons and space-time singularities. The known Schwarzschild metric is static and describes the exterior of a BH with mass $M=G\,m$ [we shall consider units with $c=1$], but without charge and angular moment. The outer solutions are well known and vastly studied. However, the inner solutions are still subject of discussion and interpretation. There are many different approaches to this solutions. The standard relativistic view is that of an empty space-time with a singularity at $r=0$. Other theories argue that, like any other BH-like objets (for example a relativistic ultra compact star) has some mass distribution\cite{BHD1}.

It is standard in the literature, to use this metric by making a Wick rotation $\tau=i\,t$ in order to study the interior of the B-H. Euclidean space-times are analytic continuations of Lorentzian ones. The complex rotation of the time coordinate to imaginary values yields many insights into the quantum nature of space and time. These methods are motivated by the formal similarity between the propagator for quantum mechanics, and the partition function in statistical mechanics. The results obtained by these methods agree in agreement with those obtained by completely different techniques, like for example the Hawking temperature and entropy of black holes (\cite{hawking}, \cite{Wald}). The corresponding space, called Euclidean black hole (EB-H), has interesting mathematical properties and important physical applications\cite{Frolov}. We can see that the near-horizon geometry for a EB-H factorizes into the product of a sphere of radius $2M$ and a $\tau-r$ structure, which looks like a plane parameterized in polar coordinates, with $\frac{\tau}{4M}$ playing the role of the angular variable. If the periodicity of that variable is $2 \pi$, the geometry will be regular; otherwise it will exhibit a conical singularity\cite{Wilczek}. The regular geometry for EB-H occurs when $\tau$ is periodic, with period $8 \pi M$. Identifying the periodicity with an inverse temperature, we are led to\footnote{Throughout this paper we use Planck units.}
\begin{equation}
\beta=8 \pi M, \qquad T_{BH} = \left(\frac{\hbar c^3}{G k}\right) \frac{1}{\beta}.
\end{equation}
This $T_{BH}$ matches the Bekenstein-Hawking (BH) temperature\cite{bek} of the B-H, which originally was calculated in a quite different way\cite{hawking}. Given that result for the temperature, it is usual to define the thermodynamic entropy, as
\begin{equation}\label{entropy}
S(M)=\int_{0}^{M}\frac{dm}{T(m)}=\left(\frac{G k}{\hbar c^3}\right) \frac{A}{4},
\end{equation}
where $A=4\pi(2M)^{2}$ is the area of the two-sphere over the horizon. \\

The Schwarzschild B-H describes a spherically symmetric compact object, the relevant coordinates to describe the B-H are $t$, and $r$.
In order to study the interior of a Schwarzschild B-H, we introduce the metric:
\begin{equation}\label{metric}
dS^{2}=-f(r)dt^{2}+\frac{dr^{2}}{f(r)}-r^{2}d\Omega^{2},
\end{equation}
where $d\Omega^2 = d\theta^2+\sin^2(\theta) \,d\phi^2$ and $f(r)=1-\frac{2M}{r}<0$ in the interior of the B-H.

\section{Boundary conditions in the minimum action principle and back-reaction effects}

It is known that in the event that a manifold has a boundary
$\partial{\cal{M}}$, the action should be supplemented by a
boundary term for the variational principle to be
well-defined\cite{1,Hawking}. However, this is not the only manner to
study this problem. As was recently demonstrated\cite{rb,rb1}, there
is another way to include the flux around a hypersurface that
encloses a physical source without the inclusion of another term
in the Einstein-Hilbert (EH) action
\begin{equation}
 S_{EH} = \frac{1}{2k} \int d^{4}x \, \sqrt{-g} \,\left[\frac{\hat{R}}{2\kappa} + \hat{{\cal L}}\right],
\label{eqn:EinsteinHilbert}
\end{equation}
by making a constraint on the variation of the EH action: $\delta S_{EH} =0$:
\begin{equation}\label{delta}
\int d^{4}x \sqrt{-\hat{g}} \left[ \delta g^{\alpha \beta} \left( \hat{G}_{\alpha \beta} + \kappa\, \hat{T}_{\alpha \beta} \right) + \hat{g}^{\alpha \beta} \delta R_{\alpha \beta} \right] =0,
\end{equation}
where $\kappa=8\pi G/c^4$, $\hat{T}_{\alpha\beta}=2\frac{\delta \hat{{\cal L}}}{\delta g^{\alpha\beta}}-\hat{g}_{\alpha\beta} \hat{{\cal L}}$ is the stress tensor that describes matter and $\hat{{\cal L}}$ is the Lagrangian density, and $\hat{G}_{\alpha \beta}=\hat{R}_{\alpha \beta}-(1/2)\,\hat{g}_{\alpha \beta}\,\hat{R}$ is the background Einstein tensor. The last term in (\ref{delta}) is very important because takes into account boundary conditions. When that quantity is zero, we obtain the well known Einstein's equations without cosmological constant. This element can be written as:
\begin{equation}\label{bor}
\hat{g}^{\alpha\beta} \delta R_{\alpha\beta} =
\left[\delta W^{\alpha}\right]_{|\alpha} - \left( \hat{g}^{\alpha\epsilon} \right)_{|\epsilon}  \,\delta\Gamma^{\beta}_{\alpha\beta} +
 ( \hat{g}^{\alpha\beta} )_{|\epsilon}  \,\delta\Gamma^{\epsilon}_{\alpha\beta},
\end{equation}
where $\hat{g}^{\alpha\beta} \delta R_{\alpha\beta} =\delta \Phi(x^{\alpha}) = {\Lambda}\, \hat{g}^{\alpha\beta} \delta
g_{\alpha\beta}$ is the flux of the 4-vector ${\delta W}^{\alpha}= {\delta\Gamma}^{\epsilon}_{\beta\epsilon} \hat{g}^{\beta\alpha}- {\delta \Gamma}^{\alpha}_{\beta\gamma} \hat{g}^{\beta\gamma}$ that crosses any $3D$ closed manifold defined on an arbitrary region of the background manifold, which is considered as Riemannian and is characterized by the Levi-Civita connections. The flux that crosses the 3D-gaussian hypersurface, $\delta\Phi$, is related to the cosmological constant and the variation of the scalar field: $\delta\sigma$:
\begin{equation}
\delta\Phi = - \frac{4}{3} \Lambda\,\delta \sigma.
\end{equation}

In order to calculate $\delta R_{\alpha \beta}$, we shall use the Palatini identity\cite{pal}
\begin{equation}
\delta{R}^{\alpha}_{\beta\gamma\alpha}=\delta{R}_{\beta\gamma}= \left(\delta\Gamma^{\alpha}_{\beta\alpha} \right)_{| \gamma} - \left(\delta\Gamma^{\alpha}_{\beta\gamma} \right)_{| \alpha}.
\end{equation}
A very important fact is that the fields $\delta \bar{W}^{\alpha}$ are invariant under gauge-transformations $\delta \bar{W}^{\alpha} = \delta W^{\alpha} - \nabla_{\alpha} \delta \Phi$, where $\delta \Phi$ satisfy $\Box \delta \Phi=0$. A very important fact is that the fields $\delta \bar{W}^{\alpha}$ are invariant under gauge-transformations $\delta \bar{W}_{\alpha} = \delta W_{\alpha} - \nabla_{\alpha} \delta \Phi$, where $\delta \Phi$ satisfy $\Box \delta \Phi=0$. Due to this fact, it is possible to define the redefined background Einstein's tensor
\begin{equation}\label{tr}
\hat{G}_{\alpha \beta} = G_{\alpha \beta} - \lambda(x) \,\hat{g}_{\alpha \beta},
\end{equation}
where
\begin{equation}
\lambda(x)\equiv \left<{\Lambda}\left(x^{\mu}\right)\right>=\oint\,d\Sigma\,\sqrt{-\hat{g}}\,\, {\Lambda}\left(x^{\mu}\right),
\end{equation}
such that $d\Sigma$ is the differential of the closed hypersurface given by the cyclic coordinates. They are coordinates on which the background metric tensor $\hat{g}_{\alpha\beta}$, are independent.  Because $\Lambda$ has a quantum origin, we must be more specific about the meaning of $\left<{\Lambda}\left(x^{\mu}\right)\right>\equiv \left<B\left|{\Lambda}\left(x^{\mu}\right)\right|B\right>$.
In our case the background quantum state can be represented in a ordinary Fock space in contrast with Loop Quantum Gravity\cite{lqg1,lqg2}, where operators are qualitatively different
from the standard quantization of gauge fields. In order to implement the formalism we must describe the variation of the connections with respect to the background manifold, which is a Riemannian one. To extend the Riemann manifold we shall consider the connections
\begin{equation}\label{ConexionWeyl}
\Gamma^{\alpha}_{\beta\gamma} = \left\{ \begin{array}{cc}  \alpha \, \\ \beta \, \gamma  \end{array} \right\} + \delta \Gamma^{\alpha}_{\beta\gamma} = \left\{ \begin{array}{cc}  \alpha \, \\ \beta \, \gamma  \end{array} \right\}+ \beta \,\sigma^{\alpha} \hat{g}_{\beta\gamma}.
\end{equation}
The last term is a geometrical displacement $\delta \Gamma^{\alpha}_{\beta\gamma}=\beta \,\sigma^{\alpha} \hat{g}_{\beta\gamma}$ with respect to the background (Riemannian) manifold, described with the Levi-Civita connections. The particular case $\beta=1/3$ guarantees the integrability of boundary terms in (\ref{bor}). The condition of integrability expresses that we can assign univocally a norm to any vector in any point, so that it must be required that $\hat{g}^{\alpha\beta} \delta R_{\alpha\beta}=\nabla_{\alpha} \delta W^{\alpha}$. Of course, this is a particular case of (\ref{bor}). In particular, the case $\hat{g}^{\alpha\beta} \delta R_{\alpha\beta}=0$, gives us the standard Einstein's equations: $ \hat{G}_{\alpha\beta}+\kappa\,\hat{T}_{\alpha\beta}=0$.

In this background must be fulfilled: $\Delta g_{\alpha \beta} = \hat{g}_{\alpha \beta ; \gamma} dx^{\gamma}=0$. However, on the extended manifold (\ref{ConexionWeyl}), we obtain
\begin{equation}\label{VariaciongWeyl}
\delta g_{\alpha \beta} = {g}_{\alpha \beta | \gamma} dx^{\gamma} = - \frac{1}{3} (\sigma_{\beta} \hat{g}_{\alpha \gamma} + \sigma_{\alpha} \hat{g}_{\beta \gamma}) dx^{\gamma},
\end{equation}
where ${g}_{\alpha \beta | \gamma}$ denotes the covariant derivative on the extended manifold, once ${g}_{\alpha \beta ; \gamma}=0$.
Therefore, the variation of the Ricci tensor on the extended manifold will be
\begin{eqnarray}
\delta R_{\alpha \beta} & = &\left( \delta \Gamma^{\epsilon}_{\alpha \epsilon} \right)_{|\beta} - ( \delta \Gamma^{\epsilon}_{\alpha \beta} )_{|\epsilon} \nonumber \\
 &=& \frac{1}{3} \left[ \nabla_{\beta} \sigma_{\alpha} + \frac{1}{3} \left( \sigma_{\alpha} \sigma_{\beta} + \sigma_{\beta} \sigma_{\alpha} \right)\right. - \left. \hat{g}_{\alpha \beta} \left( \nabla_{\epsilon} \sigma^{\epsilon} + \frac{2}{3} \sigma_{\nu} \sigma^{\nu} \right) \right], \label{VariacionRicciWeyl}
\end{eqnarray}
such that the variation of the scalar curvature is: $\delta R = \nabla_{\mu} \delta W^{\mu} = \nabla_{\mu} \sigma^{\mu} + \sigma_{\mu} \sigma^{\mu}$. Therefore, the extended Einstein tensor $\delta G_{\alpha\beta}=\delta R_{\alpha\beta} - \hat{g}_{\alpha\beta}\, \delta R$, will be
\begin{equation}\label{ein}
\delta G_{\alpha \beta}  = \frac{1}{3} \left[ \nabla_{\beta} \sigma_{\alpha} + \frac{1}{3} \left( \sigma_{\alpha} \sigma_{\beta} + \sigma_{\beta} \sigma_{\alpha} \right) + \frac{1}{2} \hat{g}_{\alpha \beta}  \nabla_{\epsilon} \sigma^{\epsilon} + \frac{1}{3} \hat{g}_{\alpha \beta} \sigma_{\nu} \sigma^{\nu} \right].
\end{equation}
Hence, if we require that $\delta G_{\alpha \beta}=-\Lambda\,\hat{g}_{\alpha\beta}$, we obtain that $\delta G=\hat{g}^{\alpha\beta}\,\delta G_{\alpha  \beta}$ is
\begin{equation}
\delta G = -4 \Lambda\left(\sigma,\sigma_{\mu}\right).
\end{equation}
The cosmological constant $\Lambda$ is an invariant on the background manifold, but not on the extended one: $\Lambda(\sigma,\sigma_{\alpha})=-\frac{1}{4} \left(\frac{2}{3} \sigma_{\alpha} \sigma^{\alpha} + \Box \sigma \right)$, on which behaves as a functional. We can define the action
\begin{equation}\label{AccionLambda}
\mathcal{W} = \int d^{4}x \sqrt{-\hat{g}} \,\Lambda(\sigma, \sigma_{\alpha}).
\end{equation}
If we require that $\delta \mathcal{W} = 0$ we obtain that $\sigma$ is a free scalar field on the extended manifold: $\Box \sigma = 0$. The scalar field $\sigma$ describes the back reaction effects which leaves invariant the action:
\begin{eqnarray}
{\cal S}_{EH} &=& \int d^4 x\, \sqrt{-\hat{g}}\, \left[\frac{\hat{R}}{2\kappa} + \hat{{\cal L}}\right] \nonumber \\
&=& \int d^4 x\, \left[\sqrt{-\hat{g}} e^{-\frac{2}{3}\sigma}\right]\,
\left\{\left[\frac{\hat{R}}{2\kappa} + \hat{{\cal L}}\right]\,e^{\frac{2}{3}\sigma}\right\},
\end{eqnarray}
and if we require that $\delta {\cal S}_{EH} =0$, we obtain
\begin{equation}
-\frac{\delta V}{V} = \frac{\delta \left[\frac{\hat{R}}{2\kappa} + \hat{{\cal L}}\right]}{\left[\frac{\hat{R}}{2\kappa} + \hat{{\cal L}}\right]}
= \frac{2}{3} \,\delta\sigma,
\end{equation}
where $\delta\sigma = \sigma_{\mu} dx^{\mu}$ is an exact differential and $V=\sqrt{-\hat{ g}}$ is the volume of the Riemannian manifold.
Furthermore, $ \Pi^{\alpha}=\frac{\delta \Lambda}{\delta \sigma_{\alpha}}=-{1\over 4} \sigma^{\alpha}$ is the geometrical momentum and the
dynamics of $\sigma$ describes a free scalar field
\begin{equation}\label{si}
{\Box} \sigma =0,
\end{equation}
so that the momentum components $\Pi^{\alpha}$ comply with
the equation
\begin{equation}
{\nabla}_{\alpha} \Pi^{\alpha} =0.
\end{equation}
If we define the scalar invariant
$\Pi^2=\Pi_{\alpha}\Pi^{\alpha}$, we obtain that
\begin{equation}
\left[\sigma,\Pi^{2}\right] = \frac{1}{16}\left\{ \sigma_{\alpha} \left[\sigma,\sigma^{\alpha} \right]
 + \left[\sigma,\sigma_{\alpha} \right] \sigma^{\alpha} \right\}=0,
\end{equation}
where we have used that $\sigma_{\alpha} \hat{U}^{\alpha} = \hat{U}_{\alpha} \sigma^{\alpha}$, and\cite{rb1}
\begin{equation}\label{con}
\left[\sigma(x),\sigma^{\alpha}(y) \right] =- i \Theta^{\alpha}\, \delta^{(4)} (x-y), \qquad \left[\sigma(x),\sigma_{\alpha}(y) \right] =
i \Theta_{\alpha}\, \delta^{(4)} (x-y),
\end{equation}
with $\Theta^{\alpha} = \hbar\, \hat{U}^{\alpha}$. Therefore we can define
the relativistic invariant $\Theta^2 = \Theta_{\alpha}
\Theta^{\alpha} = \hbar^2 \hat{U}_{\alpha}\, \hat{U}^{\alpha}$,
where $\hat{U}^{\alpha}={dx^{\alpha}\over dS}$ are the relativistic components of the Riemannian
velocities.

\section{Back-reaction in the interior of B-H and quantum thermodynamics}

\begin{figure}[h]
\noindent
\includegraphics[width=.3\textwidth]{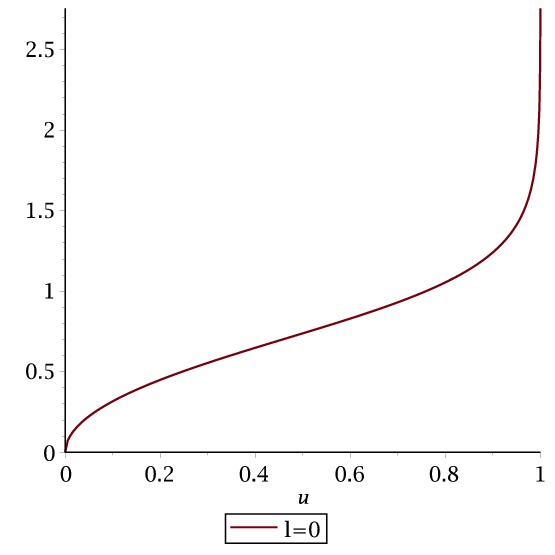}\vskip -0cm\caption{$R_{00}$ for $0<u<1$, using $C_{1}=1$ and $C_{2}=0$.}\label{F1}
\end{figure}

\begin{figure}[h]
\noindent
\includegraphics[width=.3\textwidth]{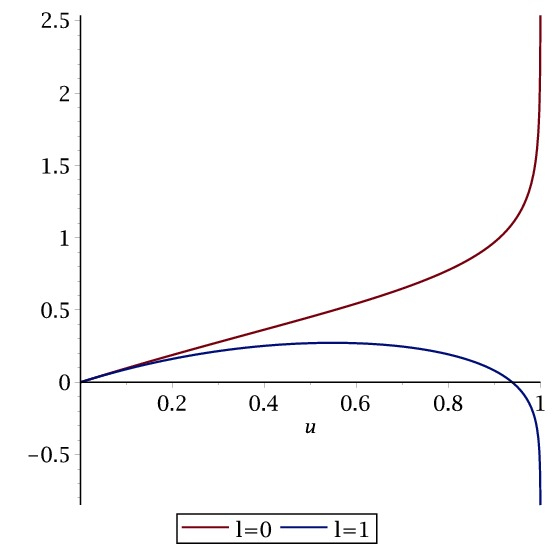}\vskip -0cm\caption{$R_{10}$ and $R_{11}$ for $0<u<1$, using $C_{1}=1$ and $C_{2}=0$.}\label{F2}
\end{figure}

\begin{figure}[h]
\noindent
\includegraphics[width=.3\textwidth]{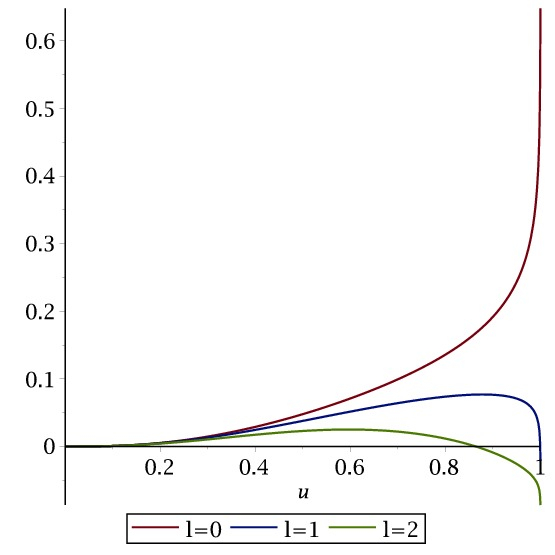}\vskip -0cm\caption{$R_{50}$, $R_{51}$ and $R_{52}$ for $0<u<1$, using $C_{1}=1$ and $C_{2}=0$.}\label{F3}
\end{figure}

The interior B-H metric with back-reaction effects included is
\begin{equation}
\begin{tiny}
g_{\mu\nu} = {\rm diag}\left[ -e^{(2/3)\sigma} f(r),  \frac{e^{-(2/3)\sigma}}{f(r)}, - e^{-(2/3)\sigma} \, r^2,-e^{-(2/3)\sigma} \, r^2 \sin^2 (\theta) \right].
\end{tiny}
\end{equation}
In order to describe the back-reaction effects in the interior of the B-H with line element (\ref{metric}), we must consider solutions of the equation $\Box\sigma=0$, for $r< 2M$. The massless scalar field $\sigma$ for the line element \eqref{metric} and $r<2M$ is described by a superposition of modes $\sigma_{klm}(r,\theta,\phi) = R_{kl}(r)\,T_{k}(t)\,Y_{lm}(\theta,\phi)$, where the functions $Y_{lm}(\theta,\phi)$ are the usual spherical harmonics. In this case, the temporal and radial equations, after take $u=1-\frac{r}{2M}>0$, are
\begin{equation}\label{Temporal}
\frac{\partial^{2}T_{n}}{\partial t^{2}}+\left(\frac{E_{n}}{\hbar}\right)^2\,T_n(t)=0,
\end{equation}
\begin{eqnarray}
\frac{\partial^{2}R_{nl}(u)}{\partial u^{2}}&=&\left[ \left( \frac{2 M E_n}{\hbar} \right)^2 \left( \frac{u-1}{u} \right)^2 + \frac{l (l+1)}{u(u-1)} \right] R_{nl}(u)\nonumber \\
&-&  \frac{(2 u -1)}{u(u-1)} \frac{\partial R_{nl}(u)}{\partial u}, \label{Radialu}
\end{eqnarray}
where the radial equation can be expressed in terms of the confluent Heun functions\cite{Muniz}\cite{Vieira}\cite{Li}\cite{Birkandan}: ${\cal H_C}$
\begin{eqnarray}
R_{nl}(u)&=& e^{{\alpha_{n}\over 2}\,u}\,\left[ C_{1}\,u^{\alpha_{n}}\,{\cal H_C}(\alpha_{n},\beta_{n},\gamma_{n},\delta_{n},\eta_{nl},u) \right. \nonumber \\
&+& \left. C_{2}\,u^{-\alpha_{n}}\,{\cal H_C}(\alpha_{n},-\beta_{n},
\gamma_{n},\delta_{n},\eta_{nl},u)\right],  \label{RadialSolution}
\end{eqnarray}
such that $0<u<1$ and the parameters $\alpha_{n}$, $\beta_{n}$, $\gamma$, $\delta_{n}$ and $\eta_{nl}$, are given by
\begin{eqnarray}
\alpha_{n}&=& 4M\,\left(\frac{E_{n}}{\hbar}\right), \quad \beta_{n}=\alpha_n,\quad \gamma_n=0, \nonumber \\ \delta_{n}&=& 8M^2\,\left(\frac{E_{n}}{\hbar}\right)^2,\quad \eta_{nl}=-8M^2\,\left(\frac{E_{n}}{\hbar}\right)^2-l(l+1). \nonumber
\end{eqnarray}
Following \cite{Fiziev}, we must impose the so called $\delta_{n}$ and $\Delta_{N+1}$ conditions, respectively
\begin{subequations}
\begin{equation}\label{dcond}
\frac{\delta_n}{\alpha_n}+\frac{\beta_n+\gamma_n}{2}+1=-n,
\end{equation}
\begin{equation}
\Delta_{N+1}=0,
\end{equation}
\end{subequations}
with $n$ a positive integer. Provided the fulfilment of the above two $\delta$-conditions, the confluent Heun solutions reduces to a polynomial of degree N\cite{Fiziev}.
The condition \eqref{dcond} for ${\cal H_C}(\alpha_{n},\beta_{n},\gamma_{n},\delta_{n},\eta_{nl},u)$ gives the following expression for the energy levels $E_{n}$:
\begin{equation}\label{elevels}
E_{n}\,r_{sch,n}=\frac{\hbar}{2},
\end{equation}
where $r_{sch,n}=2M_{n}$ and $M_{n}=\frac{M}{(n+1)}$. The expression \eqref{elevels} is very important because it tells us that the uncertainly principle is fulfilled for each energy level. A similar result, but only for $n=0$, was obtained in\cite{Villari}. We can see that for this expression that the admissible energy-levels are inversely proportional to the Schwarzschild radius\cite{Schurman}. With these results, now the parameters of the Confluent Heun function are
\begin{eqnarray}
\alpha_{n}&=&(n+1), \qquad \beta_{n}= \alpha_n, \qquad \gamma_n=0, \nonumber \\
\delta_{n}&=& \frac{1}{2} (n+1)^2, \qquad \eta_{nl}= -\frac{1}{2} (n+1)^2 -l(l+1). \nonumber
\end{eqnarray}
Finally, we can write the complete solution for the field $\sigma(t,r,\theta,\phi)$, as
\begin{equation}
\sigma(t,r,\theta,\phi)= \sum_{n=0}^{N-1}\,\sigma_n(t,r,\theta,\phi),
\end{equation}
where
\begin{eqnarray}
\sigma_n(t,r,\theta,\phi)&=& \sum_{l\geq L_{-}}^{L_{+}}\,\sum_{m =-l}^{l} \left[A_{nlm}\,\sigma_{nlm}(t,r,\theta,\phi) \right.\nonumber \\
&+&\left. A^{\dagger}_{nlm}\,\sigma^*_{nlm}(t,r,\theta,\phi)\right],
\end{eqnarray}
such that the modes $\sigma_{nlm}(t,r,\theta,\phi)$, are:
\begin{equation}
\sigma_{nlm}(t,r,\theta,\phi) = \left(\frac{E_n}{\hbar}\right)^2 \, R_{nl}(r)\, Y_{lm}(\theta,\phi)\, T_n(t),
\end{equation}
with a radial local solution expanded around $u=0$\footnote{The confluent Heun functions ${\cal H_C}(n+1,n+1,0,\frac{1}{2} (n+1)^2, -\frac{1}{2} (n+1)^2 -l(l+1),u)$ is a particular solution of the differential equation
\begin{equation}
\frac{\partial^{2}{\cal H_C}}{\partial u^{2}}+\left(n+1+\frac{n+2}{u}+\frac{1}{u-1}\right)\frac{\partial {\cal H_C}}{\partial u}+\left(\frac{\mu_{nl}}{u}+\frac{\nu_{nl}}{u-1}\right)\,{\cal H_C}=0,
\end{equation}
where $\mu_{nl}=(n+1)^{2}+l(l+1)$ and $\nu_{nl}=(n+1)-l(l+1)$.
If we demand $\nu>0$ as like the rest of parameters from Heun equations, we obtain upper and lower bounds for $l$ in function of $n$:
\begin{equation}
L_{-} \leq l \leq  L_{+},
\end{equation}
where $L_{\pm}=-\frac{1}{2} \left( 1 \mp \sqrt{5+4n} \right)$. \\
}:
\begin{small}
\begin{eqnarray}\label{rad}
R_{nl}(u)&=&C_{1}\, e^{\left(\frac{n+1}{2}\right)\,u}\,\, {\cal H_C}\left[\alpha_n,\beta_n,0,\delta_n, \eta_{nl},u\right].
\end{eqnarray}
\end{small}
In the figures (\ref{F1}), (\ref{F2}) and (\ref{F3}), we show different plots of the functions $R_{nl}$, for $n=0$, $n=2$ and $n=5$, respectively. Notice that in all the cases the state function is well behaved at $u=0$, but not at $u=1$. This is because the solution we have
founded is a local function which is valid only around $u=0$.\\

If we suppose that exist a lower bound $M_{p}=G\,m_p$ on the mass of EB-H, there is a condition for the possible values of $n$:
\begin{equation}
n \leq \frac{M}{M_{p}}-1 \leq N-1,
\end{equation}
where $N$ is the lower integer that verify $\frac{M}{M_{p}} \leq N$.

\begin{figure}[h]
\noindent
\includegraphics[width=.3\textwidth]{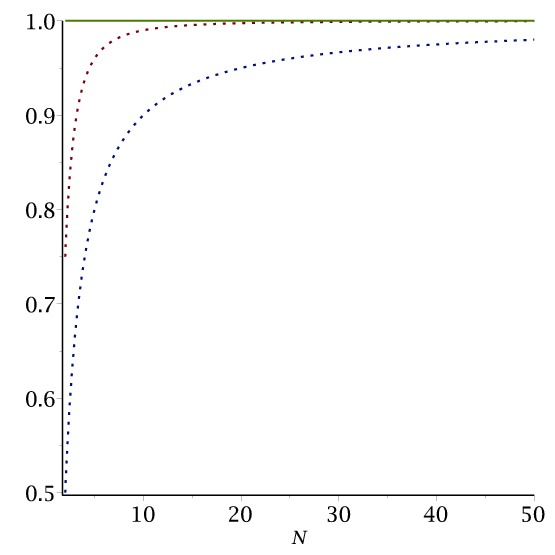}\vskip -0cm\caption{$\sum\limits_{n=0}^{N-1}\frac{\Delta A_{n}}{16 \pi M^{2}}$ (red dots) and $\sum\limits_{n=0}^{N-1}\frac{\Delta M_{n}}{M}$ (blue dots) for different values of $N$.}\label{F4}
\end{figure}

From a classical point of view, we calculate the difference in mass between two consecutive $M_{n}$ states, $\Delta M_{n}$:
\begin{equation}\label{dM}
\Delta M_{n} = M_{n} - M_{n+1} = \frac{M}{(n+1)(n+2)}
\end{equation}
The discrete values of $M_{n}$ allow us to also discretizate the area $A_{n}$ and entropy $S_{n}$, according to \eqref{entropy}. We can determine the difference between two consecutive area states $\Delta A_{n}>0$:
\begin{equation}
\Delta A_{n} = 4 \pi \left(r_{n}^{2} - r_{n+1}^{2}\right) = 16 \pi M^{2} \left( \frac{2n+3}{(n+1)^{2}(n+2)^{2}} \right).
\end{equation}
Furthermore, we can calculate the total contribution of the area, as the sum over the difference between consecutive $nth$-areas
\begin{equation}
\sum\limits_{n=0}^{N-1}\Delta A_{n} = 16 \pi M^{2} \left( 1-\frac{1}{(N+1)^{2}} \right) ,
\end{equation}
that depends on the number of states: $N$. In the same way, we obtain the entropy of the B-H, from the sum over the difference between consecutive $nth$-entropies
\begin{equation}
\begin{split}
\sum\limits_{n=0}^{N-1}\Delta S_{n} & =   \left(\frac{G k}{\hbar c^3}\right) \frac{1}{4}\,\sum\limits_{n=0}^{N-1}\Delta A_{n} \\ & = \left(\frac{G k}{\hbar c^3}\right) 4\pi M^{2} \left( 1-\frac{1}{(N+1)^{2}} \right).
\end{split}
\end{equation}
In the figure (\ref{F4}) is shown (with red dots) that $\sum\limits_{n=0}^{N-1}\frac{\Delta A_{n}}{16 \pi M^{2}}$ tends to the classical value as $N$ increases. Finally, if we calculate the temperature $T$ in the context of the first law of B-H mechanics:
\begin{equation}
T_{N}=\frac{\sum\limits_{n=0}^{N-1} \Delta M_{n}}{\sum\limits_{n=0}^{N-1} \Delta S_{n}}=\left(\frac{\hbar c^3}{G k}\right) \frac{1}{4 \pi M} \frac{1}{\left( 1 + \frac{1}{N+1} \right) },
\end{equation}
we obtain that $T_{BH} \leq T_{N} < 2\,T_{BH}$, for different values of $N$. However, this is not a true temperature, because in the interior of the B-H the states are entangled, so that $T_N$ must be interpreted as a {\it latent temperature}. It is interesting to notice that, if we have a single state, i.e., if $N=1$, we recover the BH temperature: $\frac{M_{0}}{S_{0}}=T_{BH}$, which could be related to a true temperature because this state is associated to an energy level which is at the Schwarzschild radius: $r_{sh,0}=2\,M_0=2\,M$. When the number of states is very large: $N\gg 1$, we obtain the up cut for $T_N$:
\begin{equation}
\begin{split}
\lim_{N\rightarrow \infty} T_N & = \left(\frac{\hbar c^3}{G k}\right) \lim_{N\rightarrow \infty} \frac{1}{4 \pi M} \frac{1}{\left( 1 + \frac{1}{N+1} \right) } \\ & \rightarrow \left(\frac{\hbar c^3}{G k}\right) \frac{1}{4 \pi M} =2\, T_{BH}.
\end{split}
\end{equation}
In the figure (\ref{F4}) we have plotted $\frac{1}{A_{BH}}\sum\limits_{n=0}^{N-1}\Delta A_{n}$ (red dots) and $\frac{1}{M} \sum\limits_{n=0}^{N-1}\Delta M_{n}$ (blue dots) for different values of $N$. Notice both, $\sum\limits_{n=0}^{N-1}\Delta A_{n}$
and $\sum\limits_{n=0}^{N-1}\Delta M_{n}$, approaches respectively to their BH values, $A_{BH}$ and $M$, when $N\rightarrow \infty$.

\section{Final comments}

By using the Relativistic Quantum Geometry formalism, we have calculated the quantum structure of space-time in the interior of a Schwarzschild B-H. Unfortunately, because the metric has a non-removable singularity at $r=0$, it is not possible calculate normalised solutions for $\sigma$. However, we have founded a local solution around $r=2M$, as a finite superposition of polynomials that describe the energy level of space-time. This is a very important result because we have verified that the uncertainly principle $E_n\, r_{sh,n}=\hbar/2$ is fulfilled on each energy level of the B-H. This provide us with the possibility to develop a quantum thermodynamic, where energy, mass, temperature and entropy are dependent with the number of states in the B-H. An important fact is that the B-H entropy is obtained in the limit case where the number of states tends to infinity, but the BH temperature is obtained for a B-H with an unique state, which is associated to the extreme level with a radius given by the Schwarzschild one: $r_{sh,0}=2\,M_0\equiv 2\,M$. However, in general the quantum temperature here obtained is in the range: $T_{BH} \leq T_{N} < 2\,T_{BH}$, and must be considered as a latent temperature because the quantum coherence of the system.

\section*{Acknowledgements}

\noindent The authors acknowledge CONICET, Argentina (PIP 11220150100072CO) and UNMdP (EXA852/18) for financial support.

\newpage
\end{document}